\documentclass[aps,twocolumn,showpacs]{revtex4}

\usepackage{graphicx}

\def\Rb{{\bf R}}  \def\rb{{\bf r}}  \def\Ab{{\bf A}}
\def\qb{{\bf q}}  \def\kb{{\bf k}}  \def\gb{{\bf g}}
\def\ee{{\rm e}}  \def\ii{{\rm i}}

\begin{document}

\title{Measurement of electron wave functions and confining potentials via photoemission}

\author{A. Mugarza,$^{1,2}$ J. E. Ortega,$^{1,2}$ F. J. Himpsel,$^3$ and F. J. Garc\'{\i}a de Abajo$^{1,4}$}

\affiliation{${^1}$Donostia International Physics Center (DIPC),
Aptdo. 1072, 20080 San Sebasti\'{a}n, Spain\\
       ${^2}$Departamento de F\'{\i}sica Aplicada I, Universidad del Pa\'{\i}s Vasco, Plaza O$\tilde{n}$ati 2, 20018 San Sebasti\'{a}n Spain\\
       $^{3}$Department of Physics, University of Wisconsin Madison\\
       $^{4}$Centro Mixto de Materiales CSIC/UPV, Paseo Manuel Lardizabal 3, E-20018 San Sebastian, Spain}

\date{\today}

\begin{abstract}
Wave functions and electron potentials of laterally-confined
surface states are determined experimentally by means of
photoemission from stepped Au(111) surfaces. Using an iterative
formalism borrowed from x-ray diffraction, we retrieve the
real-space wave functions from the Fourier transform of their
momentum representations, whose absolute values in turn are
directly measured by angle-resolved photoemission. The effective
confining potential is then obtained by introducing the wave
functions into Schr\"{o}dinger's equation.

\end{abstract}

\pacs{79.60.Bm,68.35.Bs,73.21.-b}

\maketitle


The electron wave functions and potentials in low-dimensional
systems are of primary importance for tailoring electronic
properties of nanostructures. The electron energy levels and the
probability density are the physical observables. These can be
obtained in real space from local conductance maps via scanning
tunneling microscopy/spectroscopy (STM/STS)
\cite{Avouris1,Buergi1}, or in reciprocal space by angle-resolved
photoemission \cite{Kawakami1,Mugarza1}. The latter has
been used to study thin films \cite{Kawakami1} or
two-dimensional arrays of nano-objects \cite{Mugarza1}, whereas
STM/STS resolves individual nanostructures on a
surface.

The question arises whether it is possible to directly derive wave
functions and/or electron potentials from experimental data. The
standard procedure consists of assuming a model potential, whose
parameters are obtained by fitting the experiment. For
instance, in core-level photoelectron diffraction muffin-tin
potentials are assumed for retrieving atomic positions that fit
the experiment \cite{KFH78_1,GVF01_1}. The electron density of
surface states confined between two steps has been measured by STM
\cite{Avouris1,Buergi1} and has been modeled by semitransparent
mirrors located at the step edges \cite{Buergi1}. It has been
argued that the position of the mirror has to be moved slightly
away from the steps to account for electron spill over
\cite{Otero1}.

We introduce a direct method to determine the effective electron
potential from angle-resolved photoemission data, without any a
priori knowledge about its nature. All the necessary information
is contained in the momentum distribution of the photoemission
intensity. This allows us to unambiguously determine the potential
and wave functions of one-dimensional quantum well states on
stepped Au(111). The real-space wave functions are derived by
Fourier transform of their momentum-space representations, the
square of which is proportional to the photoelectron intensity
under the conditions discussed below. However, the phase in
momentum space is not measured and this leads to the well-known
phase problem in optics, shared with many other
techniques, such as x-ray and electron diffraction \cite{Saldin}.
Various iteration methods have been devised to retrieve real-space
objects from the modulus of their momentum space representation,
in particular the oversampling method \cite{Miao}. These methods
are valid for a confined wave function, because the phase is
obtained by repeatedly diminishing the amplitude of the wave
function outside the confinement region. The one-electron potential can be
obtained after dividing the Schr\"odinger equation by the wave
function.

\begin{figure}
\centerline{{\includegraphics[width=\linewidth]{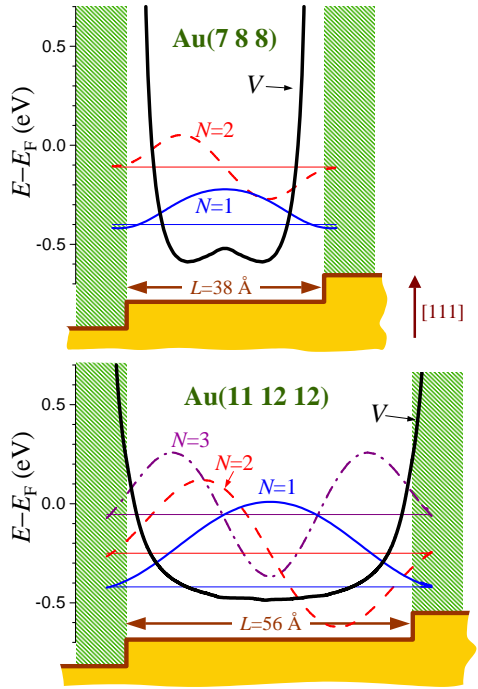}}}
\caption{Vicinal Au(111) surfaces with quantum well states
confined by step edges. The energy levels are directly measured by
angle-resolved photoemission. The confining potential $V$ and the
wave functions of states $N$=1,2,3 are obtained from the momentum
distribution of the photoemission intensity.} \label{Fig1}
\end{figure}

\begin{figure}
\centerline{{\includegraphics[width=\linewidth]{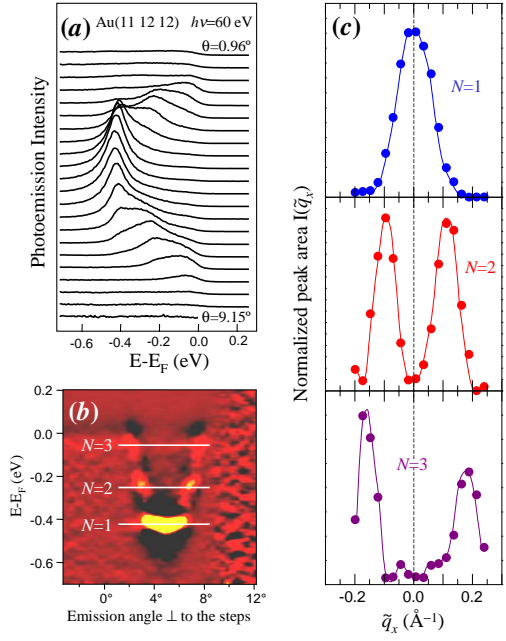}}}
\caption{{\bf (a)} Photoemission spectra from Au(11,12,12) showing
one-dimensional surface states. {\bf (b)} Second energy derivative
of the intensity showing the presence of three quantum levels
$E_N^{exp}$. {\bf (c)} Peak area of each quantum level obtained by
line fitting of individual spectra in (a). The lines are
spline-fits to the data points.} \label{Fig2}
\end{figure}

As a test system we use vicinal noble metal surfaces characterized
by equally-spaced, linear step arrays and one-dimensional surface
states confined within individual terraces of nominal size $L$.
That is the case of Au(111) vicinal surfaces with relatively wide
terraces, such as Au(788) ($L=38$ \AA) and Au(11,12,12) ($L=56$
\AA) shown in Fig.\ \ref{Fig1} \cite{Mugarza1,Mugarza2}. On
Au(11,12,12), the lowest three quantum-well levels ($N=1-3$) lie
below the Fermi energy and they can be probed by photoemission, as
shown in Figs.\ \ref{Fig2}(a)-(b). The emission angle $\theta$ is
given relative to the terrace normal in the plane perpendicular to
the step array. The data have been taken at the PGM beam line in
the Synchrotron Radiation Center (SRC) of the UW-Madison, using a
hemispherical Scienta SES200 spectrometer. The energy and angular
resolution were 20+7 meV (photons + electrons) and 0.3$^{\circ}$,
respectively. The light is p-polarized and incident 60$^{\circ}$
off-normal. By tuning the photon energy to 60 eV we avoid overlap
between surface umklapp replicas \cite{Mugarza2}. Fig.\
\ref{Fig2}(c) shows the angular photoemission intensity scan for
each quantum level as obtained from the areas of the peaks in
Fig.\ \ref{Fig2}(a). The scale for the component of the
photoelectron wave vector parallel to the terrace $q_x$ has been
determined from the emission angle and the kinetic energy $E$ as
$q_{x}=\sin \theta \sqrt{2mE}/\hbar$. The parallel momentum scale
is shifted by the reciprocal step lattice vector $g=2\pi/L=0.11$
\AA$^{-1}$ to bring it into the first Brillouin zone
\cite{Mugarza2} and corrected for the small parallel photon
momentum of $0.026$ \AA$^{-1}$.

The photoemission intensity in reciprocal space $\qb$ can be
understood in the framework of standard photoemission theory,
assuming a one-electron description. We start with the
photoemission intensity, which is proportional to the matrix
element
   \begin{eqnarray}
      I_N({\qb_\parallel}) \propto
      |<\qb|\ee^{\ii\kb\cdot\rb}\Ab\cdot\nabla|\Psi_N>|^2.
   \label{eqIN}
   \end{eqnarray}
$\Psi_N$ is the initial surface state wave function, $|\qb>$ is
the final electron state of momentum $\qb=({\qb_\parallel},q_z)$,
${\qb_\parallel}$ is the momentum component parallel to the
surface, $\Ab$ is the light polarization vector, $\kb$ is the
momentum of the photon, and $N$ refers to the quantum number of
the wave function. To proceed further, it will be assumed that the
surface state wave functions can be factorized into components
that are either parallel or perpendicular to the surface, i.e.
\cite{factorization}:
   \begin{eqnarray}
      \Psi_N(\rb) = \phi_N(\Rb) \varphi(z),
   \nonumber
   \end{eqnarray}
where $\Rb$ denotes the coordinates along the surface. If, in
addition to the surface-state confinement in the perpendicular
direction $z$, the electron is also bound to a 2D region of the
surface, the parallel component of the wave function $\phi_N(\Rb)$
must adopt a form which depends on the detailed shape of the
confining region and on the boundary conditions at its border.
Thus, $I_N$ can be written as
   \begin{eqnarray}
      I_N({\qb_\parallel}) \propto && C |<{\qb_\parallel}|\phi_N>|^2,
   \nonumber
   \end{eqnarray}
where
   \begin{eqnarray}
      C= |<q_z|\ee^{\ii k_z z}A_z \frac{\partial}{\partial z}|\varphi>|^2
   \nonumber
   \end{eqnarray}
depends very weakly on $\qb_\parallel$, so that it can be absorbed
into a normalization factor that we set to 1. Therefore, there is
a direct relation between photoemission intensity
$I_N({\qb_\parallel})$ and the Fourier transform of the wave
function $\tilde{\phi}_N({\qb_\parallel})$:
   \begin{eqnarray}
      \tilde{\phi}_N({\qb_\parallel}) = <{\qb_\parallel}|\phi_N> = \int d\Rb
      \ee^{-\ii{\qb_\parallel}\cdot\Rb}\phi_N(\Rb).
   \label{eq6}
   \end{eqnarray}
Thereby we neglect multiple scattering in the final state, which
corresponds to Fourier components
${\qb_\parallel}$+${\gb_\parallel}$ with a reciprocal lattice
vector ${\gb_\parallel}$. These affect the intensity distribution
between different Brillouin zones but not the momentum
distribution within one Brillouin zone (compare this to the spot
profiles in LEED which are not affected by multiple scattering
\cite{Henzler}). This permits obtaining the wave function in real
space $\phi_N(\Rb)$ from the intensity in reciprocal space as
 \begin{eqnarray}
      \phi_N(\Rb) = \int \frac{d{\qb_\parallel}}{(2\pi)^2} \ee^{\ii{\qb_\parallel}\cdot\Rb}
                         \sqrt{I_N({\qb_\parallel})}
                         \ee^{\ii\delta_N({\qb_\parallel})}.
   \label{eq1}
   \end{eqnarray}
However, this equation still contains an unknown phase
$\delta_N({\qb_\parallel})$. We have explored two methods to
obtain this phase: (1) an iterative procedure using oversampling
\cite{Saldin,Miao} and (2) an expansion of the wave function into
a Fourier series combined with a least squares fit. In both cases,
the strategy is minimizing $|\phi|$ outside the confinement
region.

For the iterative method, we begin with a constant phase
$\delta_N({\qb_\parallel})=0$ to start the iteration with Eq.\
(\ref{eq1}). The resulting $\phi_N(\Rb)$ is corrected outside the
confining region, and transformed back into $\qb_\parallel$ space
using Eq.\ (\ref{eq6}). The phase of
$\tilde{\phi}_N({\qb_\parallel})$ is extracted and inserted again
into Eq.\ (\ref{eq1}) to start a new iteration. The confinement
length is obtained directly from the photoemission data (see
below), and the noted correction consists of subtracting from the
newly calculated real-space wave function the wave function at the
previous step multiplied by a factor 0.1. This ensures convergence
to a wave function that vanishes outside the confinement region.
As a strong test of convergence of this iterative technique, we
have used the phase of the $N=2$ state to start the iteration of
the $N=1$ state. The first iteration step leads to a wave function
that resembles that of the $N=2$ state, a clear indication of the
importance of the phase \cite{Saldin}, and convergence to the true
$N=1$ state is achieved after several hundred iterations.

This iterative method leads to results that are in agreement with
those derived from an expansion of the wave function into sine
functions that vanish at the edges of the confinement region.
Moreover, the length of the region where the wave function takes
non-negligible values is not sensitive to the input value for the
length of the confinement region.

The size of the confining region is obtained from the Fourier
transform of the experimental intensity distribution, which can be
written as the self-convolution of the real-space wave function,
   \begin{eqnarray}
      \int \frac{d{\qb_\parallel}}{(2\pi)^2} \, \ee^{\ii{\qb_\parallel}\cdot\Rb} I_N({\qb_\parallel}) =
      \int d\Rb' \phi_N(\Rb') \phi_N^*(\Rb'-\Rb).
   \label{eq4}
   \end{eqnarray}
This convolution takes non-zero values in an area twice the size
of the confining region. For the data of Fig.\ \ref{Fig2} in
particular, one obtains curves that are basically confined within
a region of $\approx 60$ \AA in diameter (not shown), in agreement
with the nominal terrace width in Au(11 12 12), $L=56$ \AA.

The natural normalization for both the two-dimensional wave
function $\phi_N$ and the measured intensity is provided by Eq.\
(\ref{eq4}) if one sets $\Rb=0$:
   \begin{eqnarray}
      \int \frac{d{\qb_\parallel}}{(2\pi)^2} I_N({\qb_\parallel}) = <\phi_N|\phi_N> = 1.
   \nonumber
   \end{eqnarray}
Applying this procedure to the angular scans of Fig.\
\ref{Fig2}(c), we have obtained the surface state wave functions
for quantum-well levels in Au(11 12 12) that are shown in Fig.\
\ref{Fig1}. All three wave functions are confined to a region
whose width matches the terrace width $L$. Thus the wave functions
exhibit a clear terrace confinement that has not been assumed by
our reconstruction procedure, but rather it has emerged from the
information contained in the photoemission data.


The actual effective potential of the terrace can be retrieved
from the Schr\"{o}dinger equation as
   \begin{eqnarray}
      V(\Rb)-E_N = \frac{\hbar^2}{2 m^* \phi_N(\Rb)} \nabla^2\phi_N(\Rb),
   \label{eq2}
   \end{eqnarray}
where $m^*=0.26 m$ is the effective mass of the electron in the
initial state of energy $E_N$, taken from the dispersion of the
surface state on flat Au(111). The potential becomes independent
of the quantum number $N$ if many-electron effects are absent, as
we assumed initially.

\begin{figure}
\centerline{{\includegraphics[width=\linewidth]{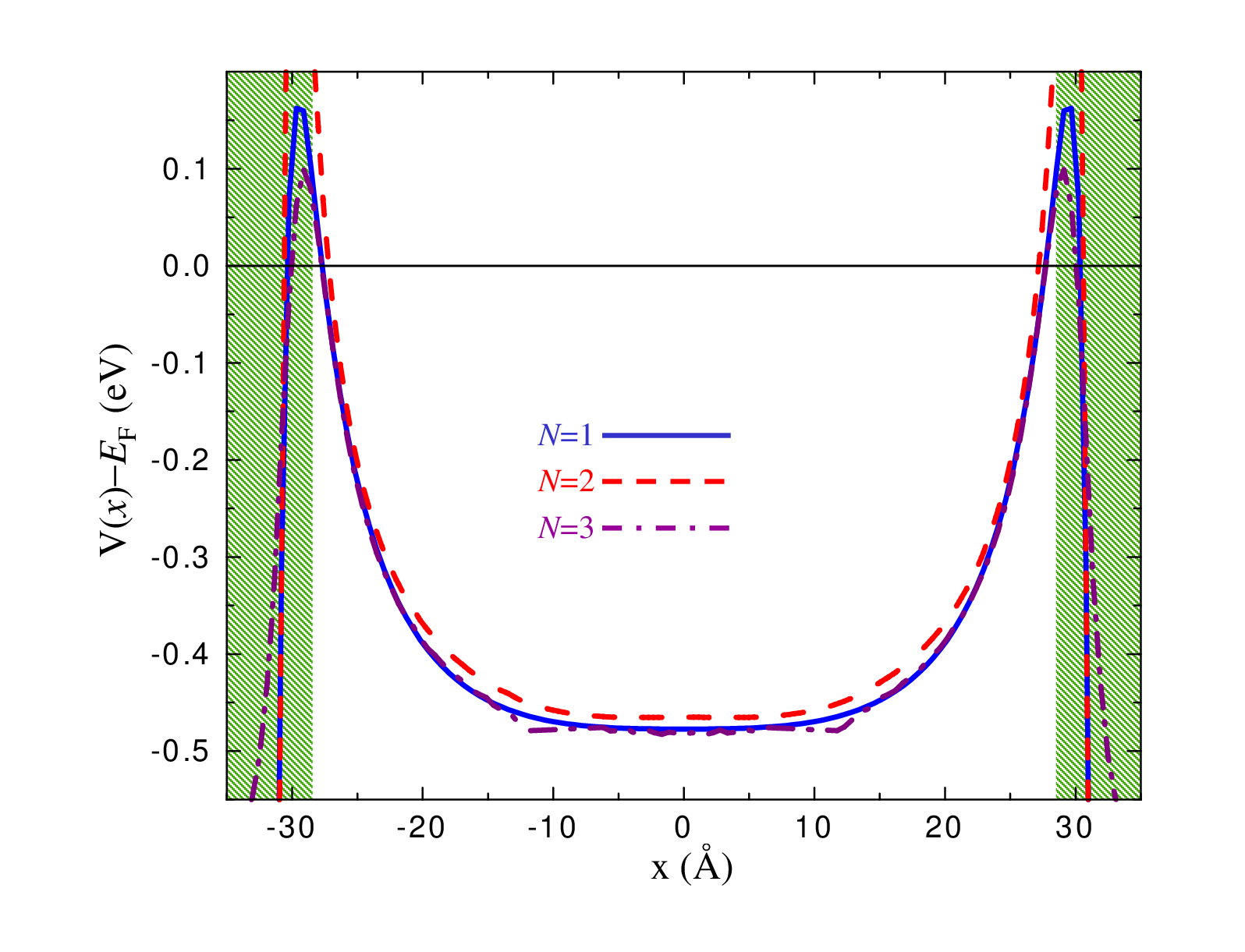}}}
\caption{Confining potentials $V$ obtained from the experimental
photoemission intensities of Fig.\ \ref{Fig2} for the $N$=1,2,3
states in Au(11,12,12) using the oversampling method.}
\label{Fig3}
\end{figure}

The experimental wave functions of Fig.\ \ref{Fig1} have been
introduced into Eq.\ (\ref{eq2}) and the resulting electron
potential has been represented in Fig.\ \ref{Fig3}. In order to
compare the potential derived from the quantum states with various $N$
we add on the left side of Eq.\ (\ref{eq2}) the respective
experimental values $E_N^{\rm exp}$ shown in Fig.\ \ref{Fig1}
(horizontal bars). Excellent mutual agreement is obtained for the
shape of the potential derived independently from each of the
three wave functions $N=1-3$, proving the validity of our method.
The potential exhibits a smooth central region and sharp
boundaries that force electron confinement. It must be stressed
that the validity of Eq.\ (\ref{eq2}) is limited to regions where
the wave function is not too small, and therefore, the asymptotic
limit of the step barrier potential cannot be determined.

There is a slight uphill/downhill asymmetry of $\phi$ and $V$,
which is related to an asymmetry of $|\tilde{\phi}|$ for $\pm q$ .
This effect shows up for the highest level ($N=3$), which is the
least confined. Such an asymmetry explains the asymmetric
reflectivity observed in STM \cite{Buergi1} where the downhill
step reflects more strongly than the uphill. The asymmetry
increases for higher-lying empty states which start spilling over
the weaker potential barrier on the uphill side.

The same iterative procedure has been applied to the
one-dimensional quantum well of Au(788), measured in a different
system \cite{Mugarza1}. In this case the terraces are 38 \AA wide
and only the first two quantum levels are occupied. We also obtain
a good consistency in the electron potential between the two
levels. The resulting average is displayed in Fig.\ \ref{Fig1},
and compared with the average potential of Au(11,12,12). In
contrast to the latter case, the width of the potential well in
Au(788) is narrower than the terrace size, as also observed in
thin Pb films \cite{Otero1}.


In summary, we have introduced a simple procedure for directly
obtaining wave functions and effective potentials of confined
electronic states from the momentum distribution of the
photoemission intensity. This procedure, which relies on
photoemission theory within the one-electron approach, has been
successfully tested in the case of laterally confined surface
states on Au(111) vicinal surfaces. The method is of general
applicability for discrete states because they are spatially
confined and allow an iterative determination of the phase by
forcing $\phi$=0 outside the confinement region. It can be applied
to any type of surface nanostructure confined in at least one
dimension, such as arrays of quantum wires and quantum dots. While
STM provides direct images of the low spatial frequencies,
photoemission determines the high spatial frequencies. It would be
interesting to pursue the reverse transformation from the real
space $|\phi|$ in STM to $\tilde{\phi}$ in $q$ space using the
technique described here. A connection between $|\phi|$ and the
$E(q)$ band dispersion has been made already \cite{Buergi1,Horn}


A.Mu. and J.E.O. are supported by the Universidad del Pa\'{\i}s
Vasco (9/UPV 00057.240-13668/2001) and the Max Planck Research
Award Program. J.G.deA. acknowledges support by the Spanish
Ministerio de Ciencia y Tecnolog\'{\i}a. F.J.H. acknowledges
support by the NSF under Award Nos. DMR-9815416 and DMR-0084402.




\begin{thebibliography}{99}
\bibitem{factorization} The approximation of a factorized wave
function is reasonable for states that are confined to a single
terrace island. It breaks down for a superlattice with strong
overlap between adjacent terraces where $\phi$ delocalized and
discrete states become bands.
\end{thebibliography}
\end{document}